\documentstyle[eqsecnum,aps]{revtex}
\begin{document}

\begin{titlepage}
\baselineskip .15in
\begin{flushright}
WU-AP/72/98
\end{flushright}

\begin{center}
{\bf

\vskip 1cm
{\large Equivalence of black hole thermodynamics 

between a generalized theory of gravity and the Einstein theory}

}\vskip .6in

{\sc  Jun-ichirou Koga}${}^{*}$\\[1em]

{\em Department of Physics, Waseda University,
Shinjuku-ku, Tokyo 169-8555, Japan} \\[3em] 
  
{\sc  Kei-ichi Maeda}${}^{\dagger}$\\[1em]

{\em Department of Physics, Waseda University,
Shinjuku-ku, Tokyo 169-8555, Japan} \\[1em]
{\em and  Astronomy Centre, University of Sussex, Brighton BN1 
9QJ, U.K.}

\end{center}
\vfill
\begin{abstract}
We analyze black hole thermodynamics in a generalized theory of 
gravity whose Lagrangian is an arbitrary function of the metric, 
the Ricci tensor and a scalar field. We can convert the theory 
into the Einstein frame via a ``Legendre" transformation or a 
conformal transformation. We calculate thermodynamical variables 
both  in the original frame and in the Einstein frame, following 
the Iyer--Wald definition which satisfies the first law of 
thermodynamics. We show that all thermodynamical variables defined 
in the original frame are the same as those in the Einstein 
frame, if the spacetimes in both frames are asymptotically flat, 
regular and possess event horizons with non-zero temperatures. 
This result may be useful to study whether the second law is 
still valid in the generalized theory of gravity.
\end{abstract}

\vskip 1cm
\begin{center}
March, 1998
\end{center}
\vfill
\baselineskip .2in
$*$~electronic mail : koga@gravity.phys.waseda.ac.jp\\
$\dagger$~electronic mail : maeda@gravity.phys.waseda.ac.jp\\
\end{titlepage}

\normalsize

\renewcommand{\large}{\normalsize}
\renewcommand{\Large}{\normalsize}
\renewcommand{\huge}{\normalsize}
\section{Introduction}

Black hole thermodynamics was originally formulated within Einstein's 
theory of general relativity, by showing that 
analogous relations to the usual thermodynamical laws hold in black 
hole dynamics\cite{BCH}. 
After Hawking's discovery\cite{Hawking} of black hole evaporation due 
to black body radiation with the temperature, $T = \hbar \kappa / 2 
\pi$, where $\kappa$ is the surface gravity of the black hole, black 
holes have been regarded as real thermodynamical objects. While 
Hawking's discovery has elucidated the thermodynamical nature of 
black holes, it also left problems which have not been solved 
yet, such as the information loss problem, fates of black holes due to 
the Hawking evaporation and the origin of the black hole 
entropy. Those problems are expected to be explained by quantum 
gravity. Actually, a recent development in string theory, which is one 
of the most 
promising candidates of quantum gravity, has unveiled that 
the number of states of a D-brane configuration is exactly 
the same as the entropy of the 
corresponding black hole\cite{Dbrane}. However, such a coincidence is 
true only for 
extreme black holes because of their BPS nature. While non-extreme 
black holes are also discussed in this context (see, e.g., 
Ref.\cite{Dbrane2}), further development 
of string theory would be required to understand the origin of the 
entropy of the generic black holes.

Together with string theory, investigations based on field theory 
including stringy effects may also be helpful to understand such 
microscopic black holes and their properties. One possible method to 
study the stringy effects is to consider effective field theories,  in 
which 
corrections to Einstein gravity are expected\cite{Callan}, e.g., 
higher curvature 
interactions and a dilaton coupling. Higher curvature corrections arise 
also in the effective theory with quantum loop effects\cite{BDBook}. 
Those effective theories may not be 
fundamental, and the corrections should be regarded as 
perturbative effects. However, such corrections may modify black hole 
physics from that in the Einstein theory, which includes one of the 
most important properties 
of microscopic black holes. In addition, by 
investigating the generalized theories of gravity, as the effective 
theories, we may extract a universal 
feature of black hole 
physics, which cannot be understood by the analysis within the 
Einstein theory. 

Since Hawking's result that a black hole is regarded as a 
thermodynamical object does not 
depend on the details of the Einstein equations, we expect that a 
black hole in the generalized theories of gravity also shows thermal 
properties. In addition, when the horizon of the black hole is 
bifurcate, the zero-th law was proved\cite{RaczWald} and the first law 
has been formulated in an arbitrary diffeomorphism invariant 
theory\cite{Wald,IyerWald}. Therefore, we expect that the 
thermodynamical 
laws 
of black holes may also hold in the generalized theories. However, the 
entropy in dynamical processes is still to be defined, and thus, not only 
the third law, but also the second law has not been established so 
far, in general. For the theories which can be transformed into the 
Einstein frame via a conformal transformation, the entropy is proved to 
be the same as that in the Einstein frame, and the second law is proved 
for some specific models, by examining the null energy condition in 
the Einstein frame\cite{JKM1}. 
Equivalence between the entropy of a stationary black hole in the 
original frame and that in the 
Einstein frame is also shown in the theory whose Lagrangian 
includes a specific form of higher curvature interactions\cite{JKM2}, 
based 
on the fact that the transformation into the Einstein frame does not 
alter the asymptotic structure of the spacetime, such as the mass and 
the angular momentum of the black hole. It is not the case, however, 
when the theory contains a scalar field which couples non-minimally to 
gravity.

In this paper, we generalize those results into a much wider class of 
theories. The Lagrangian we consider is an arbitrary function of the 
metric, the Ricci tensor, a scalar field and its derivative. This class of 
theories includes any combinations of the 
Ricci tensor of any order as well as a non-minimal coupling of a scalar 
field. We convert the theory into the Einstein frame via a ``Legendre" 
transformation\cite{Magnano} or a conformal 
transformation\cite{conftrans}, and show that all the thermodynamical 
variables are the same between the original frame and the Einstein 
frame. It is expected that the equivalence of the thermodynamical 
variables may provide a useful clue to the second law, as in the case 
of Ref.\cite{JKM1}. 

In the formulation of the first law in Ref.\cite{Wald,IyerWald}, the 
assumption that a black hole has a bifurcate 
Killing horizon plays a crucial role. This assumption seems to be 
reasonable if the event horizon is a Killing horizon, since a Killing 
horizon with non-vanishing temperature should be 
bifurcate\cite{RaczWald}, and the third law implies that a zero 
temperature state cannot be realized in physical 
processes, although its general proof is still absent. 
We then assume, in 
this paper, that the event horizon is a Killing horizon, and focus only on 
black holes with non-zero temperature.

As we mentioned above, we consider the effective theories, which may 
not be adequate to study physics at the Planck scale. Hence, we will be 
concerned only with black holes sufficiently larger than Planck scale. 
The corrections of the higher curvature interactions in 
those cases will not be large enough to change drastically the geometry 
of a black hole spacetime from that in the Einstein theory. However, 
they might change black hole thermodynamics. This is the present 
subject.  

We organize this paper as follows. We first describe the transformation 
of the theory into the Einstein frame in the next section. In 
Section III, we briefly repeat the definition of the thermodynamical 
variables 
of a black hole in the generalized theory of gravity formulated in 
Ref.\cite{Wald,IyerWald}. By following their definition, we show,  in 
Section IV, the equivalence between the thermodynamical variables in 
the original frame and those in the Einstein frame. The conclusion and 
discussions are described in 
Section V. The Appendix contains derivations of the equations used in 
our 
calculations. 

We use two metrics $g^{\mu \nu}$ and $\overline{g}^{\mu 
\nu}$ for fundamental variables\cite{note1}, and then, the inverse 
matrices of $g^{\mu \nu}$ and 
$\overline{g}^{\mu \nu}$ will be denoted as 
$g^{ - 1}_{\mu \nu}$ and $\overline{g}^{ - 1}_{\mu \nu}$, respectively, 
because two variables, $\overline{g}_{\mu \nu}^{ - 1}$ and 
$g_{\mu \alpha}^{ - 1} g_{\nu \beta}^{ - 1} \overline{g}^{\alpha \beta}$ 
(or $g_{\mu \nu}^{ - 1}$ and 
$\overline{g}_{\mu \alpha}^{ - 1} \overline{g}_{\nu \beta}^{ - 1} 
 g^{\alpha \beta}$), must be distinguished, although both of them are 
the same and are expressed as $\overline{g}_{\mu \nu}$ (or 
$g_{\mu \nu}$) in the standard convention. The units we adopt are 
$c = G = k_{\mbox{\scriptsize B}} = 1$. 

\section{Transformation of theory}

The action we consider is written as
\begin{equation}
I = \frac{1}{16 \pi} \int d^4 x \: \sqrt{- g} \; f \! \left( g^{\alpha \beta}, 
R_{\alpha \beta}, \phi, \nabla_{\alpha} \phi \right) \; ,
\label{eqn:actionorign}
\end{equation}
where $f$ is an arbitrary function of the metric, 
$g^{\alpha \beta}$, the Ricci tensor, $R_{\alpha \beta}$, a scalar field, 
$\phi$, and 
its derivative. $f$ of course includes an arbitrary function of a scalar 
curvature, $R$, because $R = g^{\alpha \beta} R_{\alpha \beta}$. We 
transform the action(\ref{eqn:actionorign}) into the 
Einstein frame by defining a new ``metric'' tensor, $\overline{g}^{\mu 
\nu}$, as 
\begin{equation}
\overline{g}^{\mu \nu} \equiv \frac{\partial (\sqrt{- g} f)}{\partial 
R_{\mu \nu}} \frac{1}{\sqrt{- \det \left( \partial (\sqrt{- g} 
f) / \partial R_{\alpha \beta} \right)}} = \frac{\sqrt{- g}}{\sqrt{- 
\overline{g}}} 
\frac{\partial f}{\partial R_{\mu \nu}} \; .
\label{eqn:gbardef}
\end{equation}
Since $\overline{g}^{\mu \nu}$ is a function of $g^{\alpha \beta}$, 
$R_{\alpha \beta}$, $\phi$ and $\nabla_{\alpha} \phi$, it is written as 
\begin{equation}
\overline{g}^{\mu \nu} = 
\overline{g}^{\mu \nu}(g^{\alpha \beta}, R_{\alpha \beta}, \phi, 
\nabla_{\alpha} \phi) \; .
\label{eqn:gbarrelation}
\end{equation}
For an arbitrary function $f$, there are two cases, depending on 
whether or not $\det | \partial^2 f / \partial R_{\mu \nu} \partial 
R_{\rho \sigma} |$ vanishes\cite{NoteDet}.

If $\det | \partial^2 f / \partial R_{\mu \nu} \partial 
R_{\rho \sigma} | = 0$ (the degenerate case), we may use a conformal 
transformation to convert the theory into the Einstein frame. For 
example, if the action is given by
\begin{equation}
I = \frac{1}{16 \pi} \int d^4 x \sqrt{- g} \left[ f(\phi) R - \frac{\epsilon
(\phi)}{2} ( \nabla_{\mu} \phi) (\nabla^{\mu} \phi) - V(\phi) \right] \; ,
\label{eqn:simpleaction}
\end{equation}
we can convert the action into the Einstein 
frame\cite{conftrans} 
\begin{equation}
\overline{I} = \frac{1}{16 \pi} \int d^4 x \sqrt{- \overline{g}} \left[ 
\overline{R} 
- \frac{1}{2} \overline{g}^{\mu \nu} (\overline{\nabla}_{\mu} \Phi) 
(\overline{\nabla}_{\nu} \Phi) - U(\Phi) \right] \; ,
\label{eqn:newactionconf}
\end{equation}
via a conformal transformation, 
\begin{equation}
\overline{g}^{\mu \nu} = f^{- 1}(\phi) g^{\mu \nu} \; ,
\label{eqn:confgbar}
\end{equation}
with a redefinition of the scalar field
\begin{equation}
\Phi = \int d \phi \left[ \frac{\epsilon(\phi) f(\phi) + 3 \left[ d f(\phi) 
/ d \phi \right]^2}{f(\phi)^2} \right]^{1 / 2} \; ,
\label{eqn:saclarredef}
\end{equation}
where $U(\Phi) \equiv f^{- 2}(\phi) V(\phi)$.  

More difficulty appears when
\begin{equation}
\det \: \frac{\partial^2 f}{\partial R_{\mu \nu} \partial R_{\rho 
\sigma}} 
\neq 0 \: .
\label{eqn:invertible}
\end{equation}
In this case, we can solve Eq.(\ref{eqn:gbarrelation}) for  $R_{\mu \nu}$ 
as 
\begin{equation}
R_{\mu \nu} = 
R_{\mu \nu}(g^{\alpha \beta}, \overline{g}^{\alpha \beta}, \phi, 
\nabla_{\alpha} \phi) \; ,
\label{eqn:Riccisolved}
\end{equation}
and then, define a new action, $\overline{I}$, via a kind of Legendre 
transformation, by  
\begin{eqnarray}
\overline{I} & \equiv & \frac{1}{16 \pi} \int d^4 x \: 
\sqrt{- \overline{g}}  \: \Biggl[ \overline{g}^{\mu \nu} R_{\mu \nu}(
g^{\alpha \beta}, \partial_{\gamma} 
g^{\alpha \beta}, \partial_{\gamma} \partial_{\lambda} 
g^{\alpha \beta}) - 
\overline{g}^{\mu \nu} R_{\mu \nu}(g^{\alpha \beta}, 
\overline{g}^{\alpha \beta}, \phi, \nabla_{\gamma} \phi) 
\Biggr. \nonumber \\
& & \left. + \frac{\sqrt{- g}}{\sqrt{- \overline{g}}} \: f(g^{\alpha \beta}, 
R_{\alpha \beta}(g^{\rho \sigma}, \overline{g}^{\rho \sigma}, \phi, 
\nabla_{\rho} \phi), \phi, \nabla_{\alpha} \phi) \right] \; ,
\label{eqn:newactiondef}
\end{eqnarray}
where the independent variables are $g^{\mu \nu}$, $\overline{g}^{\mu 
\nu}$ and $\phi$. The Ricci tensor in the first term of the right hand 
side of Eq.(\ref{eqn:newactiondef}) is given by the metric, $g^{\alpha 
\beta}$, and its 
derivatives as usual, whereas those in the second and the third terms 
are given by the relation Eq.(\ref{eqn:Riccisolved}). We find, by varying 
the action(\ref{eqn:newactiondef}), the Euler--Lagrange equation for 
$\overline{g}^{\mu \nu}$ gives the same relation as Eq.(\ref
{eqn:gbardef}), and the total variations of $I$ and $\overline{I}$ are 
found to be the 
same when we substitute Eq.(\ref{eqn:gbardef}). 
Eq.(\ref{eqn:newactiondef}) is rewritten into the form 
\begin{eqnarray}
\overline{I} & = & \frac{1}{16 \pi} \int d^4 x \: \sqrt{- \overline{g}} 
\: \Biggl[ 
\overline{R}(\overline{g}^{\alpha \beta}, \partial_{\gamma} 
\overline{g}^{\alpha \beta}, \partial_{\gamma} \partial_{\lambda} 
\overline{g}^{\alpha \beta}) + \overline{g}^{\mu \nu} \left( 
\overline{\Gamma}^{\rho}_{\rho \sigma} 
\overline{\Gamma}^{\sigma}_{\mu \nu} - 
\overline{\Gamma}^{\rho}_{\mu \sigma} 
\overline{\Gamma}^{\sigma}_{\rho \nu} \right) \Biggr. \nonumber \\
& & \left. - \overline{g}^{\mu \nu} R_{\mu \nu}(g^{\alpha \beta}, 
\overline{g}^{\alpha \beta}, \phi, \overline{\nabla}_{\alpha} \phi) + 
\frac{\sqrt{- g}}{
\sqrt{- \overline{g}}} f(g^{\alpha \beta}, R_{\alpha \beta}(g^{\rho 
\sigma}, \overline{g}^{\rho \sigma}, \phi, \overline{\nabla}_{\rho} 
\phi), \phi, \overline{\nabla}_{\alpha} \phi) \right]  
\: ,
\label{eqn:actionEinstein}
\end{eqnarray}
by using the relation
\begin{eqnarray}
R_{\mu \nu}(g^{\alpha \beta}, \partial_{\gamma} g^{\alpha \beta}, 
\partial_{\gamma} \partial_{\lambda} g^{\alpha \beta}) & = & 
\overline{R}_{\mu \nu}(\overline{g}^{\alpha \beta}, \partial_{\gamma} 
\overline{g}^{\alpha \beta}, 
\partial_{\gamma} \partial_{\lambda} \overline{g}^{\alpha \beta}) 
\nonumber \\
& & + \: \overline{\Gamma}^{\rho}_{\rho \sigma} 
\overline{\Gamma}^{\sigma}_{\mu \nu} - 
\overline{\Gamma}^{\rho}_{\mu \sigma} 
\overline{\Gamma}^{\sigma}_{\nu \rho} + \overline{\nabla}_{\rho} 
\overline{\Gamma}^{\rho}_{\mu \nu} - \overline{\nabla}_{\mu} 
\overline{\Gamma}^{\rho}_{\rho \nu} \; ,
\end{eqnarray}
where $\overline{\Gamma}^{\rho}_{\mu \nu} \equiv \frac{1}{2} 
g^{\rho \sigma}
\left( \overline{\nabla}_{\mu} g^{-1}_{\nu \sigma} + 
\overline{\nabla}_{\nu} g^{-1}_{\mu \sigma} - 
\overline{\nabla}_{\sigma} g^{-1}_{\mu \nu} \right)
$, $\overline{\nabla}_{\nu}$ denotes the covariant derivative 
associated with $\overline{g}^{\mu \nu}$, and  the total divergence is 
omitted in Eq.(\ref{eqn:actionEinstein})\cite{NoteTotDiv}. 
Now $\overline{g}^{\mu \nu}$ plays the role of the metric, and the first 
term in the right hand side of Eq.(\ref{eqn:actionEinstein}) is the 
Einstein--Hilbert action of $\overline{g}^{\mu \nu}$. The second term 
is interpreted as a kinetic term of an exotic spin-2 matter field, 
$g^{\mu \nu}$, since it is bilinear in the first derivatives of 
$g^{\mu \nu}$. The 
rest consists of a kinetic term of the scalar field, $\phi$, potential and 
interaction of the matter fields. Hence $\overline{I}$ is regarded as the 
action of Einstein gravity with exotic matter, which provides the 
equivalent equations of motion to those derived from the original 
action, $I$. The equations of motion in the Einstein frame are 
second order differential equations, while those in the original frame 
are fourth order.

Although we considered the degenerate case separately, the conformal 
transformation(\ref{eqn:confgbar}) is consistent with 
Eq.(\ref{eqn:gbardef}), and the action(\ref{eqn:newactionconf}) is 
formally obtained by substituting Eq.(\ref{eqn:confgbar}) and 
(\ref{eqn:saclarredef}) into Eq.(\ref{eqn:actionEinstein}), where the 
Ricci tensor terms given by Eq.(\ref{eqn:Riccisolved}) cancel. Since we 
need only Eq.(\ref{eqn:gbardef}) and Eq.(\ref{eqn:actionEinstein}) to 
show the equivalence of the thermodynamical variables, we can treat 
the action(\ref{eqn:simpleaction}) on an equal footing, as the 
general case, Eq.(\ref{eqn:actionorign}). 

However, if Eq.(\ref{eqn:invertible}) is not satisfied and the action 
is not 
written as Eq.(\ref{eqn:simpleaction}), a transformation into the 
Einstein frame is not known. We will not discuss such cases in 
what follows.

\section{Definition of thermodynamical variables}

The thermodynamical variables of black holes in the generalized 
theories of gravity are defined such that they satisfy the first law of 
black hole thermodynamics\cite{Wald,IyerWald}.  In Ref.\cite
{IyerWald}, the first law is established for an arbitrary perturbation 
from a regular background spacetime of a black hole, for which the 
following assumptions are made: (1) It is stationary and axisymmetric, 
and hence it possesses two Killing vectors, $\xi^{\mu}_{(t)}$ and 
$\xi^{\mu}_{(\varphi)}$, associated with the respective symmetry. 
(2) It 
is asymptotically flat, and then the mass and the angular momentum 
are well-defined. (3) The event horizon of the black hole is a 
bifurcate Killing horizon. 
Definition of the thermodynamical variables is based on the covariance 
of the Lagrangian density 4-form, $\epsilon_{\mu \nu \rho \sigma} L$, 
under a 
diffeomorphism, where $L$ is a scalar function of the field variables 
which we denote as $\psi_{(i)}$ collectively. The variation of 
$\epsilon_{\mu \nu \rho \sigma} L$ is described as 
\begin{equation}
\delta \left( \epsilon_{\mu \nu \rho \sigma} L \right) = \epsilon_{\mu 
\nu \rho \sigma} E^{(i)} \delta \psi_{(i)} + \epsilon_{\mu \nu \rho 
\sigma} \nabla_{\beta} \Theta^{\beta}[\psi_{(i)}, 
\delta \psi_{(i)}]  \; ,
\label{eqn:variation}
\end{equation}
where $E^{(i)} = 0$ is the equation of motion for $\psi_{(i)}$, and 
$\Theta^{\beta}[\psi_{(i)}, \delta \psi_{(i)}]$ in the totally 
divergent term is a functional of the field variables, $\psi_{(i)}$, and 
their variations, $\delta \psi_{(i)}$. When we identify the variation, 
$\delta \psi_{(i)}$, with a general coordinate transformation, ${\cal 
L}_{\zeta} \psi_{(i)}$, induced by an arbitrary vector field, 
$\zeta^{\mu}$, we obtain
\begin{equation}
\nabla_{\beta} \left( \zeta^{\beta} L \right) = - E^{(i)} \: 
{\cal L}_{\zeta} \psi_{(i)} + \nabla_{\beta} \Theta^{\beta}[\psi_{(i)}, 
{\cal L}_{\zeta} \psi_{(i)}] \; .
\end{equation}
Then the vector field, 
$J^{\beta}[\psi_{(i)}, \zeta^{\rho}, {\cal L}_{\zeta} \psi_{(i)}] 
\equiv 
\Theta^{\beta}[\psi_{(i)}, {\cal L}_{\zeta} \psi_{(i)}] - 
\zeta^{\beta} L
$, is divergence-free, $\nabla_{\beta} J^{\beta} = 0$, when the 
equations of motion, $E^{(i)} = 0$, are satisfied. In addition, we can find 
an antisymmetric potential,  
$Q^{\beta \alpha}[\psi_{(i)}, \zeta^{\rho}, \nabla_{\lambda} 
\zeta^{\rho}]$, of the current, $J^{\beta}[\psi_{(i)}, \zeta^{\rho}, {\cal 
L}_{\zeta} \psi_{(i)}]$, such that $J^{\beta} = \nabla_{\alpha} 
Q^{\beta \alpha}$\cite{Wald2}. $\Theta^{\beta}[\psi_{(i)}, 
\delta \psi_{(i)}]$ 
and $Q^{\beta \alpha}[\psi_{(i)}, 
\zeta^{\rho}, \nabla_{\lambda} 
\zeta^{\rho}]$ are calculated in a straightforward manner for a given 
action, and those for the actions, Eq.(\ref{eqn:actionorign}) and 
Eq.(\ref{eqn:actionEinstein}), are given in the Appendix.

When we consider perturbations of the field 
variables, $\delta \psi_{(i)}$, which satisfy the linearized equations of 
motion, and set 
$\zeta^{\mu}$ to be the Killing vector, $\xi^{\mu} \equiv \xi_{(t)}^{\mu}
 +  \Omega_{\mbox{\scriptsize H}} \xi_{(\varphi)}^{\mu}$, where 
$\Omega_{\mbox{\scriptsize H}}$ is the angular velocity of the black 
hole, the symplectic current 3-form, $\omega_{\mu \nu \rho}$, which 
is given 
by the exterior derivative of 
\begin{equation}
\delta \! \left( \frac{1}{2} \epsilon_{\beta \alpha \mu \nu} Q^{\beta 
\alpha}[\psi_{(i)}, \xi^{\rho}, 
\nabla_{\lambda} \xi^{\rho}] \right) + \epsilon_{\beta \alpha \mu \nu} 
\xi^{[ \beta} \Theta^{\alpha ]}[\psi_{(i)}, \delta \psi_{(i)}] \; , 
\label{eqn:potofomega}
\end{equation}
identically vanishes. Thus, integrating $\omega_{\mu \nu \rho}$ 
over a Cauchy surface which connects the bifurcation surface of the 
Killing horizon 
with spatial infinity, we find the relation
\begin{eqnarray}
& & \delta \! \int_{\mbox{\scriptsize H}} \frac{1}{2} \epsilon_{\beta 
\alpha \mu \nu} Q^{\beta \alpha}[\psi_{(i)}, \xi^{\rho}, 
\nabla_{\lambda} \xi^{\rho}] \: 
dx^{\mu} dx^{\nu} \nonumber \\ & = & \delta \! \int_{\infty} \frac{1}{2} 
\epsilon_{\beta \alpha \mu \nu} Q^{\beta \alpha}[\psi_{(i)}, 
\xi^{\rho}, \nabla_{\lambda} \xi^{\rho}] \: dx^{\mu} dx^{\nu} + 
\int_{\infty} \epsilon_{\beta \alpha \mu \nu} \xi^{[ \beta} 
\Theta^{\alpha ]}[\psi_{(i)}, \delta \psi_{(i)}] \: dx^{\mu} dx^{\nu} 
\; .
\label{eqn:Conservation1}
\end{eqnarray}
In the integral over the bifurcation surface, H, i.e., the left hand side of 
Eq.(\ref{eqn:Conservation1}), the term proportional to $\xi^{\mu}$ 
vanishes, because $\xi^{\mu} = 0$ on that surface. If 
there exists a vector, $B^{\mu}[
\psi_{(i)}]$, whose variation gives the second term in the right hand 
side of Eq.(\ref{eqn:Conservation1}) as 
\begin{equation}
\delta \! \int_{\infty} \epsilon_{\beta \alpha \mu \nu} \xi^{[ \beta} 
B^{\alpha ]}[\psi_{(i)}] \: dx^{\mu} dx^{\nu} = \int_{\infty} 
\epsilon_{\beta \alpha \mu \nu} \xi^{[ \beta} \Theta^{\alpha ]}
[\psi_{(i)}, 
\delta \psi_{(i)}] \: dx^{\mu} dx^{\nu} \; ,
\end{equation}
Eq.(\ref{eqn:Conservation1}) provides the relation between the 
variation of 
an integral of a geometrical quantity on the horizon and that over the 
2-surface at infinity, i.e., 
\begin{eqnarray}
& & \delta \! \int_{\mbox{\scriptsize H}} \frac{1}{2} \epsilon_{\beta 
\alpha \mu \nu} Q^{\beta \alpha}[\psi_{(i)}, \xi^{\rho}, 
\nabla_{\lambda} \xi^{\rho}] \: 
dx^{\mu} dx^{\nu} \nonumber \\ & = & \delta \! \int_{\infty} \frac{1}{2} 
\epsilon_{\beta \alpha \mu \nu} \left( Q^{\beta \alpha}[\psi_{(i)}, 
\xi^{\rho}, \nabla_{\lambda} \xi^{\rho}] +  2 \xi^{[ \beta}  
B^{\alpha ]}[\psi_{(i)}] \right) 
dx^{\mu} dx^{\nu}
\; . \label{eqn:Conservation2}
\end{eqnarray}

Then, we define the entropy, $S$, the mass, $M$, and the angular 
momentum, $J$,  
of the black hole as 
\begin{eqnarray}
\frac{\hbar \kappa}{2 \pi} S & \equiv & \int_{\mbox{\scriptsize H}} 
\frac{1}{2} \epsilon_{\beta \alpha \mu \nu} 
Q^{\beta \alpha}[\psi_{(i)}, \xi^{\rho}, \nabla_{\lambda} \xi^{\rho}] 
\: dx^{\mu} dx^{\nu} \; ,
\label{eqn:Sdef} \\
M & \equiv & \int_{\infty} \frac{1}{2} \epsilon_{\beta \alpha \mu \nu} 
\left( Q^{\beta \alpha}[\psi_{(i)}, \xi_{(t)}^{\rho}, \nabla_{\lambda} 
\xi_{(t)}^{\rho}] + 2 \xi_{(t)}^{[ \beta} B^{\alpha ]}[\psi_{i}] \right) 
dx^{\mu} dx^{\nu} \; ,
\label{eqn:Mdef} \\
J & \equiv & - \int_{\infty} \frac{1}{2} \epsilon_{\beta \alpha \mu 
\nu} \left( Q^{\beta \alpha}[\psi_{(i)}, \xi_{(\varphi)}^{\rho}, 
\nabla_{\lambda} \xi_{(\varphi)}^{\rho}] + 2 \xi_{(\varphi)}^{[ \beta} 
B^{\alpha ]}[\psi_{i}] \right) dx^{\mu} dx^{\nu}
\label{eqn:Jbdef} \; ,
\end{eqnarray}
respectively, where $\kappa$ is the surface gravity of the black hole. 
Although the second term in the right hand side of Eq.(\ref{eqn:Jbdef}) 
does not contribute, we keep this term for convenience in our 
discussion below. In the case of the Einstein theory, Eq.(\ref{eqn:Sdef}) 
gives $S = A / 4 \hbar$, where $A$ is the area of the black hole, and 
Eqs.(\ref{eqn:Mdef}) and (\ref{eqn:Jbdef}) coincide with the expressions 
of the ADM mass and 
the Komar angular momentum, respectively. The existence 
of $B^{\mu}$ is closely related to asymptotic flatness. In 
Ref.\cite{IyerWald}, it is argued that the condition on the asymptotic 
behavior of the field variables should be imposed to ensure the 
existence of 
$B^{\mu}$, because the mass and the angular momentum are naturally 
defined in an asymptotically flat spacetime. 

The first law of thermodynamics, for arbitrary perturbations from a 
stationary background solution, 
\begin{equation}
\frac{\hbar \kappa}{2 \pi} \delta S = \delta M - \Omega_{
\mbox{\scriptsize H}} 
\: \delta J 
\end{equation}
follows by substituting Eqs.(\ref{eqn:Sdef}), (\ref{eqn:Mdef}) and 
(\ref{eqn:Jbdef}) into Eq.(\ref{eqn:Conservation2}), and noting that 
$Q^{\beta \alpha}[\psi_{(i)}, \zeta^{\rho}, \nabla_{\lambda} 
\zeta^{\rho}]$ is linear in $\zeta^{\rho}$ and 
$\nabla_{\lambda} \zeta^{\rho}$. 

\section{Equivalence of thermodynamical variables}

Now, following the definition in the previous section, we 
show the equivalence of the thermodynamical variables 
between the original frame and the Einstein frame, where the actions 
are given by Eqs.(\ref{eqn:actionorign}) and (\ref{eqn:actionEinstein}), 
respectively. First we have to mention further assumptions in addition 
to those in the previous section. Then, after showing the equivalence of 
the temperature and the angular velocity, we prove that the entropy, 
the mass and the angular momentum are the same in these two frames. 

\subsection{Assumptions}

Since we do not specify our model explicitly, the 
spacetime in the Einstein frame may not be regular or asymptotically 
flat, in which case we have to give up discussing black hole 
thermodynamics. 
However, when we consider the effective string theory, as an example, 
$\overline{g}^{\mu \nu}$ is typically given as 
\begin{equation}
\overline{g}^{\mu \nu} = e^{- \alpha \phi} \left( g^{\mu \nu} + \beta \: 
l_{\mbox{\tiny PL}}^2 \: g^{\mu \nu} R + \gamma \: 
l_{\mbox{\tiny PL}}^2 \: R^{\mu \nu} + \cdots \right) \; ,
\label{eqn:typicalgbar}
\end{equation}
where dots denote the higher order terms of curvature, $\phi$ is a 
dilaton field, 
$l_{\mbox{\tiny PL}}$ is the Planck length, and $\alpha$, $\beta$, 
$\gamma$ are numerical constants of order unity. 

In this case, $\overline{g}^{\mu 
\nu}$ approaches $g^{\mu \nu}$ in the asymptotically flat region in the 
original frame, if the scalar field is 
not singular, and hence the spacetime in the Einstein frame 
is also asymptotically flat. This is naturally expected to be the case in 
general, because the dominant contribution to the Lagrangian, $f$, in 
the original action(\ref{eqn:actionorign}) 
will be proportional to the scalar curvature in the asymptotically flat 
region, where $g^{\mu \nu} \rightarrow \eta^{\mu \nu} + {\cal O}(r^{- 
1})$. Although a non-minimal coupling of 
the scalar field may result in $\overline{g}^{\mu \nu} \rightarrow 
e^{\omega} 
g^{\mu \nu} + {\cal O}(r^{- 1})$, where $\omega$ is determined by the 
asymptotic value of the scalar field, we can set $\omega = 0$ by 
rescaling the length unit, without loss of generality. In what follows, 
we assume that this treatment has been done, and so $\overline{g}^{\mu 
\nu} 
\rightarrow g^{\mu \nu} + {\cal O}(r^{- 1})$. We emphasize that this 
does not necessarily indicate that 
the mass and the angular momentum are the same between the original 
frame and the Einstein frame, in contrast to Ref.\cite{JKM2}. 

In addition, if the spacetime and the scalar field in the 
original frame are regular at least outside the horizon, 
$\overline{g}^{\mu \nu}$ is also expected to be 
regular in the above example(\ref{eqn:typicalgbar}) if the black hole is 
much larger than 
Planck scale. This is because 
$\overline{g}^{\mu \nu}$ is expressed in terms of 
the field variables in the original frame as Eq.(\ref{eqn:typicalgbar}), 
and the corrections from the higher curvature terms will be 
small enough so that a singularity will not appear. This seems 
to be true in the general case for large black holes. Therefore 
we will be 
concerned only with such cases. Since 
the spacetime is regular and $\overline{g}^{\mu \nu}$ approaches 
$\eta^{\mu \nu}$ in the asymptotically flat region, the 
signature of the metric, 
$\overline{g}^{\mu \nu}$, in the Einstein frame is still Lorentzian. 

As we mentioned in the Introduction, we focus only on black holes with 
non-zero temperature. Since physics is described in the original frame, 
it might not be physically reasonable to make an assumption about the 
temperature in the Einstein frame. However, if the temperature in the 
original frame is non-zero and that in the Einstein frame vanishes, we 
naively presume that the 
black hole in the original frame is quantum mechanically unstable due 
to Hawking radiation while it might be stable in the Einstein frame. 
This 
does not seem to be natural. 
We assume, in what follows, the temperatures of black holes in both 
frames are non-zero. 

Our assumptions are summarized as: (1) $\overline{g}^{\mu \nu} 
\rightarrow g^{\mu \nu} + {\cal O}(r^{- 1})$ in the asymptotically flat 
region, where $g^{\mu \nu} \rightarrow \eta^{\mu \nu} + {\cal O}(r^{- 
1})$. (2) The transformation into the Einstein frame is regular in the 
sense that the spacetime region in the Einstein frame is regular if the 
corresponding region in the original frame is regular. (3) The 
temperatures of black holes in both frames are non-zero. 

\subsection{Temperature and angular velocity}

Here we show the equality of the temperature and the angular velocity 
in two frames\cite{note2}. For this purpose, we first show the 
Killing horizon in the Einstein frame appears at the same place as in 
the original frame. 

On the Killing horizon in the original frame, we introduce a coordinate 
system $(\theta, \varphi)$ on the 2-dimensional hypersurface 
orthogonal to $\xi^{\mu}$, such that $\xi_{(\varphi)}^{\mu} \: ( \partial 
/ \partial x^{\mu} ) \equiv 
\partial / \partial \varphi$ and denote the basis vector of the 
coordinate $\theta$ by $\theta^{\mu} \, ( \partial / \partial x^{\mu} ) 
\equiv \partial / \partial \theta$. A 2-dimensional hypersurface in the 
original frame is transformed into a 2-dimensional hypersurface in the 
Einstein frame, because the transformation is regular. We consider the 
vector fields in the Einstein frame, which are the same as $\xi^{\mu}$, 
$\xi^{\mu}_{(\varphi)}$ and $\theta^{\mu}$ in the original frame. Then 
we find that fixed points of 
$\xi^{\mu}$ in the Einstein frame appear at the same points as the 
bifurcation surface in the 
original frame, and we have, at those points,  
\begin{equation}
\overline{g}^{\: -1}_{\mu \nu} 
\xi^{\mu} \xi^{\nu} = 0 \; , ~~~~ \overline{g}^{\: -1}_{\mu \nu} \xi^{\mu} 
\theta^{\nu} = 0 \; , ~~~~ \overline{g}^{\: -1}_{\mu \nu} \xi^{\mu} 
\xi_{(\varphi)}^{\nu} = 0 \; . 
\label{eqn:normal}
\end{equation}
In addition, we notice that the Killing vectors in the 
original frame are also 
Killing vectors in the Einstein frame, because the field variables in the 
Einstein frame are composed of those in the original frame, and hence 
the Lie derivatives of the field variables in the Einstein frame with 
respect to the Killing vectors vanish. Then we have 
\begin{equation}
{\cal L}_{\xi} \left( 
\overline{g}^{\: -1}_{\mu \nu} \xi^{\mu} \xi^{\nu}
 \right) = 0 \; , ~~~~ {\cal L}_{\xi} \left( \overline{g}^{\: -1}_{\mu \nu} 
\xi^{\mu} \theta^{\nu} \right) = 0 \; , ~~~~ {\cal L}_{\xi} \left( 
\overline{g}^{\: -1}_{\mu \nu} \xi^{\mu} \xi_{(\varphi)}^{\nu} \right) 
= 0 \; , 
\end{equation}
because $\xi^{\mu}$ commutes with $\theta^{\mu}$ and 
$\xi_{(\varphi)}^{\mu}$. Therefore, we find that Eq.(\ref{eqn:normal}) 
holds along the orbits of 
$\xi^{\mu}$ through 
the fixed points of $\xi^{\mu}$, and then, 
$\xi^{\mu}$ is orthogonal to the 2-dimensional hypersurface spanned by 
$\theta^{\mu}$ and 
$\xi_{(\varphi)}^{\mu}$, is null and hence 
hypersurface orthogonal. Thus,  
$\xi^{\mu}$ 
generates a Killing horizon and the fixed points of $\xi^{\mu}$ form a 
bifurcation surface also in the Einstein frame. Since $\xi^{\mu}$ 
and the location of the bifurcation surface are the same, we see the 
horizon in the Einstein frame appears at the same place as in the 
original frame. 

There might exist other Killing horizons outside the above mentioned 
Killing horizon in the Einstein frame. However, we can show that those 
extra horizons cannot be 
bifurcate, by repeating the same argument as above. That is, 
if there exist bifurcate horizons in the Einstein frame, the 
corresponding bifurcate horizons must exist also in the original 
frame, whereas it is not the case. Hence all extra horizons must  
have vanishing temperature\cite{RaczWald}. Since we assume the 
temperatures in both frames are non-zero, we exclude such an 
exceptional case, if any, from 
our consideration.

We consider the surface gravity in order to show the equivalence of the 
temperature, since the temperature, $T$, of a black hole is given as $T 
= \hbar \kappa / 2 \pi$ in arbitrary theories of gravity. The surface 
gravity, $\overline{\kappa}$, in the Einstein frame is 
calculated, from hypersurface orthogonality of $\xi^{\mu}$, by 
\begin{equation}
2 \overline{\kappa}^2 = \left( \overline{\nabla}_{\mu} 
\xi^{\nu} \right) \left( \overline{\nabla}_{\nu} \xi^{\mu} \right) \; ,
\label{eqn:kappabar}
\end{equation}
as is the surface gravity, 
$\kappa$, in the original frame by 
\begin{equation}
2 \kappa^2 = \left( \nabla_{\mu} 
\xi^{\nu} \right) \left( \nabla_{\nu} \xi^{\mu} \right) \; . 
\label{eqn:kappa}
\end{equation}
Then, by describing $\overline{\nabla}_{\mu} \xi^{\nu} = 
\nabla_{\mu} 
\xi^{\nu} + \xi^{\rho} \Gamma^{\nu}_{\mu \rho}$, where $
\Gamma^{\rho}_{\mu \nu} \equiv \frac{1}{2} \overline{g}^{\rho \sigma} 
\left( \nabla_{\mu} \overline{g}^{\: -1}_{\nu \sigma} + 
\nabla_{\nu} \overline{g}^{\: -1}_{\mu \sigma} - 
\nabla_{\sigma} \overline{g}^{\: -1}_{\mu \nu} \right)
$, and noticing $\xi^{\mu} = 0$ at the bifurcation surface, we find  
$\overline{\nabla}_{\mu} \xi^{\nu} = \nabla_{\mu} \xi^{\nu}$, and hence, 
$\overline{\kappa} = 
\kappa$ at the bifurcation surface.  
In addition, the zero-th law, i.e., constancy of $\overline
{\kappa}$ over the horizon, holds whenever the Killing horizon is 
bifurcate\cite{RaczWald}. Therefore, $\overline{\kappa} = \kappa$, not 
only on the bifurcation surface, but all over the horizon, and therefore 
the temperatures are the same in both frames. 

The angular velocity, $\overline{\Omega}_{\mbox{\scriptsize H}}$, in 
the 
Einstein frame is given by 
\begin{equation}
\overline{g}^{\: -1}_{\mu \nu} \left( \xi_{(t)}^{\mu} + 
\overline{\Omega}_{\mbox{\scriptsize H}} \: \xi_{(\varphi)}^{\mu} 
\right) 
\xi_{(\varphi)}^{\nu} = 0 \; , 
\label{eqn:Omegabar}
\end{equation}
and we have 
\begin{equation}
\overline{g}^{\: -1}_{\mu \nu} 
\left( \xi_{(t)}^{\mu} + \Omega_{\mbox{\scriptsize H}} \: 
\xi_{(\varphi)}^{\mu} \right) \xi_{(\varphi)}^{\nu} = 0 \; ,
\label{eqn:Omega}
\end{equation}
from orthogonality of $\xi^{\mu}$ to the bifurcation surface in the 
Einstein frame. 
Hence we see that the angular velocities in the two frames are 
the same, 
$\overline{\Omega}_{\mbox{\scriptsize H}} = 
\Omega_{\mbox{\scriptsize H}}$, from Eq.(\ref{eqn:Omegabar}) and 
(\ref{eqn:Omega}).

\subsection{Entropy, mass and angular momentum}

We prove the equivalence of the entropy, the mass and the angular 
momentum, showing that the integrand of Eq.(\ref{eqn:Sdef}) on the 
bifurcation surface, 
and the asymptotic forms of the integrands of Eqs.(\ref{eqn:Mdef}) and 
(\ref{eqn:Jbdef}) at infinity are the same between the two frames. 

At the bifurcation surface ($\xi^{\mu} = 0$), the integrands of 
Eq.(\ref{eqn:Sdef}) are calculated, by setting $\zeta^{\mu} = \xi^{\mu}$ 
in 
Eqs.(\ref{eqn:AQO}) and (\ref{eqn:AQE}) and using the Killing equation 
for 
$\xi^{\mu}$, 
\begin{equation}
\frac{1}{2} \epsilon_{\beta \alpha \mu \nu} 
Q^{\beta \alpha} = - \frac{1}{16 \pi} \epsilon_{\alpha \beta \mu \nu} 
P^{\alpha \gamma} 
\nabla_{\gamma} \xi^{\beta} \; ,
\label{eqn:QcalO}
\end{equation}
in the original frame, where 
\begin{equation}
P^{\mu \nu} \equiv \frac{\partial f}{\partial R_{\mu \nu}} 
\label{eqn:Pdef} \; ,
\end{equation}
and 
\begin{equation}
\frac{1}{2} \overline{\epsilon}_{\beta \alpha \mu \nu} 
\overline{Q}^{\beta \alpha} = - \frac{1}{16 \pi} 
\overline{\epsilon}_{\alpha \beta \mu \nu} 
\overline{g}^{\alpha \gamma} \overline{\nabla}_{\gamma} \xi^{\beta} 
\; ,
\label{eqn:QcalE}
\end{equation}
in the Einstein frame, respectively. Substituting the definition of 
$\overline{g}^{\mu \nu}$, Eq.(\ref{eqn:gbardef}), 
into Eq.(\ref{eqn:QcalE}) and noticing again that 
$\overline{\nabla}_{\mu} \xi^{\nu} = \nabla_{\mu} \xi^{\nu}$ at the 
bifurcation surface, we find the tensor components of 
Eq.(\ref{eqn:QcalE}) are equal to those of Eq.(\ref{eqn:QcalO}) on that 
surface. Then, 
since $\overline{\kappa} = \kappa$ which does not vanish and the 
bifurcation surfaces in both frames appear at the same place, the 
entropy is the 
same in these two frames.  

If the integrands of Eq.(\ref{eqn:Mdef}) and Eq.(\ref{eqn:Jbdef}) are the 
same up to the order of $r^{-2}$ in the asymptotically flat 
region, the mass and the angular momentum are also the same in the 
two 
frames. Since $\overline{\epsilon}_{\alpha \beta \mu \nu}$ approaches 
$\epsilon_{\alpha \beta \mu \nu}$ as $r \rightarrow \infty$, and those 
integrands in both frames fall as $r^{-2}$, it is sufficient to show 
that 
\begin{equation}
Q^{\beta \alpha} + 2 \chi^{[ \beta} B^{\alpha ]} = 
\overline{Q}^{\beta \alpha} + 2 \chi^{[ \beta} 
\overline{B}^{\alpha ]} + {\cal O}(r^{- 3}) \; , 
\label{eqn:IntegrandInf}
\end{equation}
where $\chi^{\mu}$ is taken to be $\xi_{(t)}^{\mu}$ and 
$\xi_{(\varphi)}^{\mu}$ to show the equivalence of the mass and the 
angular momentum, respectively. 

The asymptotic forms of $Q^{\beta \alpha}$ and 
$\Theta^{\beta}$ in the original frame are calculated, from 
Eqs.(\ref{eqn:AQO}) and 
(\ref{eqn:AThetaO}), as
\begin{eqnarray}
Q^{\beta \alpha} & \rightarrow & \frac{1}{8 \pi} \left[ \nabla^{[ \alpha} 
\chi^{\beta ]} + \chi_{\nu} \nabla^{[ \beta} P^{\alpha ] \nu} + 
\chi^{[ \alpha} \nabla_{\nu} P^{\beta ] \nu} \right] + {\cal O}(r^{-3}) \; , 
\label{eqn:QasymO} \\ 
\Theta^{\beta} & \rightarrow & \frac{1}{16 \pi} \left[ 
g^{\: -1}_{\mu \nu} 
\nabla^{\beta} \delta g^{\mu \nu} - \nabla_{\nu} \delta 
g^{\nu \beta} + v^{\beta} \delta \phi \right] + {\cal O}(r^{-3}) 
\nonumber \\
& = & \frac{1}{16 \pi} \left[ g^{\: -1}_{\mu \nu} g^{\rho \beta} 
\partial_{\rho} \delta g^{\mu \nu} - \partial_{\nu} \delta g^{\nu 
\beta} + v^{\beta} \delta \phi \right] + 
{\cal O}(r^{-3}) \; , 
\label{eqn:ThetaasymO}
\end{eqnarray}
to the order of $r^{-2}$,
by replacing $\zeta^{\mu}$ with $\chi^{\mu}$ and using the facts that 
$\nabla_{\mu} \chi^{\nu} \sim {\cal O}(r^{-2})$, $P^{\mu \nu} 
\rightarrow g^{\mu \nu} + {\cal O}(r^{-1})$, $\delta g^{\mu \nu} \sim 
{\cal O}(r^{-1})$ and hence 
$\nabla_{\rho} \delta g^{\mu \nu} = \partial_{\rho} \delta g^{\mu \nu} 
+ {\cal O}(r^{-3})$, where
\begin{equation}
v^{\beta} \equiv \frac{\partial f}{\partial (\nabla_{\beta} \phi)} \; .
\label{eqn:vdef}
\end{equation} 
If there exists a vector field $C^{\beta}$ such that
$\delta C^{\beta}$ approaches $v^{\beta} \delta \phi$ and is of order 
$r^{- 2}$ at most in the asymptotically flat region, then $\Theta^{\beta} 
\rightarrow \delta B^{\beta} + {\cal O}(r^{- 3})$ as $r \rightarrow 
\infty$, where
\begin{equation}
B^{\beta} = \frac{1}{16 \pi} \left[ g^{\: -1}_{\mu \nu} g^{\rho \beta} 
\partial_{\rho} g^{\mu \nu} - \partial_{\nu} g^{\nu \beta} + 
C^{\beta} \right] \; .
\label{eqn:BO}
\end{equation}
The vector $C^{\beta}$ is expected to exist in the asymptotically flat 
spacetime, 
so that $B^{\beta}$ exists, and thus the mass and the angular 
momentum are well-defined. In the usual cases, where the kinetic term 
of $\phi$ takes the form of $(\nabla_{\mu} \phi) (\nabla^{\mu} \phi)$, 
$v^{\beta} \delta \phi$ is of order $r^{- 3}$ if $\phi$ approaches a 
constant in the asymptotically flat region, and then $C^{\beta}$ 
does not contribute to $B^{\beta}$ in those cases. 

Since $\overline{g}^{\mu \nu} \rightarrow g^{\mu 
\nu} + {\cal O}(r^{- 1})$, $\overline{Q}^{\beta \alpha}$ in the Einstein 
frame given by Eq.(\ref{eqn:AQE}) with $\zeta^{\mu} = \chi^{\mu}$ is 
calculated as 
\begin{eqnarray}
\overline{Q}^{\beta \alpha} & \rightarrow & \frac{1}{8 \pi} \biggl[ 
\overline{\nabla}^{[ \alpha} 
\chi^{\beta ]} + \chi^{\nu} H_{\mu ~ \: \nu}^{~ [ \alpha} 
\overline{g}^{\beta ] \mu} + \chi^{\nu} H_{\mu}^{~ [ \alpha \beta ]} 
\overline{g}^{\mu}_{~ \nu} + \chi^{\nu} H_{\nu \mu}^{~~ [ \alpha} 
\overline{g}^{\beta ] \mu} \biggr] + {\cal O}(r^{-3}) \nonumber \\
& = & \frac{1}{8 \pi} \overline{\nabla}^{[ \mu} \chi^{\beta ]} + 
{\cal O}(r^{-3}) \; ,
\label{eqn:QasymEo}
\end{eqnarray}
in the asymptotically flat region, where 
\begin{equation}
H_{\mu \nu}^{~~\rho} \equiv \frac{\partial \overline{f}}{\partial 
(\overline{\nabla}_{\rho} g^{\mu \nu})} \; ,
\label{eqn:Hdef} 
\end{equation}
with $\overline{f}$ being the Lagrangian of the matter fields in the 
Einstein frame, i.e., 
\begin{eqnarray}
\overline{f} & \equiv & \overline{g}^{\mu \nu} \left( 
\overline{\Gamma}_{\rho \sigma}^{\rho} 
\overline{\Gamma}_{\mu \nu}^{\sigma} - 
\overline{\Gamma}_{\mu \sigma}^{\rho}
\overline{\Gamma}_{\rho \nu}^{\sigma} \right) 
- \overline{g}^{\mu \nu} R_{\mu \nu}(g^{\alpha \beta}, 
\overline{g}^{\alpha \beta}, \phi, \overline{\nabla}_{\alpha} \phi) 
\nonumber \\
& & + \: \frac{\sqrt{- g}}{\sqrt{- \overline{g}}} \: f(g^{\alpha \beta}, 
R_{\alpha \beta}(g^{\rho \sigma}, \overline{g}^{\rho \sigma}, \phi, 
\overline{\nabla}_{\rho} \phi), \phi, \overline{\nabla}_{\alpha} \phi) 
\; , 
\label{eqn:fbardef}
\end{eqnarray}
and the symmetry of $H_{\mu \nu}^{~~\rho}$ in the first two indices, 
$H_{\mu \nu}^{~~\rho} = H_{\nu \mu}^{~~\rho}$, as well as 
$H_{\mu \nu}^{~~ \rho} \sim {\cal O}(r^{- 2})$, is used in 
Eq.(\ref{eqn:QasymEo}). 
We rewrite Eq.(\ref{eqn:QasymEo}), in 
terms of the tensors in the original frame, 
by using the relations
\begin{equation}
\overline{\nabla}_{\nu} \chi^{\beta} = \nabla_{\nu} \chi^{\beta} - 
\frac{1}{2} \chi^{\alpha} \left( \overline{g}^{\: -1}_{\nu \rho} 
\nabla_{\alpha} 
\overline{g}^{\rho \beta} + \overline{g}^{\: -1}_{\alpha \rho} 
\nabla_{\nu} 
\overline{g}^{\rho \beta} - \overline{g}^{\beta \gamma} 
\overline{g}^{\: -1}_{\alpha \rho} \overline{g}^{\: -1}_{\nu \sigma} 
\nabla_{\gamma} 
\overline{g}^{\rho \sigma} \right) \; ,
\end{equation}
and 
\begin{equation}
\overline{g}^{\mu \nu} = \frac{1}{\sqrt{- g}} \frac{1}{\sqrt{- \det 
\left( P^{\alpha \beta} \right)}} \: P^{\mu \nu} \; ,
\label{eqn:gbardef2}
\end{equation}
which follows from Eqs.(\ref{eqn:gbardef}) and (\ref{eqn:Pdef}), which 
leads to  
\begin{eqnarray}
\overline{Q}^{\beta \alpha} & = & \frac{1}{8 \pi} \left[ \nabla^{[ \alpha} 
\chi^{\beta ]} + \chi_{\nu} \nabla^{[ \beta} P^{\alpha ] \nu} + 
\frac{1}{2} \chi^{[ \beta} \nabla^{\alpha ]} P \right] + {\cal O}(r^{-3}) 
\nonumber \\
& = & Q^{\beta \alpha} + \frac{1}{8 \pi} \left[ \chi^{[ \beta} 
\nabla_{\nu} P^{\alpha ] \nu} 
+ \frac{1}{2} \chi^{[ \beta} \nabla^{\alpha ]} P \right] + 
{\cal O}(r^{-3}) \; ,
\label{eqn:QasymE}
\end{eqnarray}
where $P \equiv g_{\mu \nu} P^{\mu \nu}$, and we used 
Eq.(\ref{eqn:QasymO}). By noting that   
$\delta \overline{g}^{\mu \nu} \sim {\cal O}(r^{-1})$ and hence 
$\overline{\nabla}_{\rho} \delta \overline{g}^{\mu \nu} = 
\partial_{\rho} \delta \overline{g}^{\mu \nu} + {\cal O}(r^{-3})$ in 
the asymptotically flat region, the asymptotic form of 
$\overline{\Theta}^{\beta}$ is given, from Eq.(\ref{eqn:AThetaE}), by
\begin{eqnarray}
\overline{\Theta}^{\beta} & \rightarrow & \frac{1}{16 \pi} \left[ 
\overline{g}^{\: -1}_{\mu \nu} 
\overline{\nabla}^{\beta} \delta \overline{g}^{\mu \nu} - 
\overline{\nabla}_{\nu} \delta \overline{g}^{\nu \beta} + 
\overline{v}^{\beta} \delta \phi \right] + {\cal O}(r^{-3}) \nonumber \\
& = & \frac{1}{16 \pi} \left[ g^{\: -1}_{\mu \nu} g^{\rho \beta} 
\partial_{\rho} \delta 
\overline{g}^{\mu \nu} - \partial_{\nu} \delta \overline{g}^{\nu \beta} 
+ \overline{v}^{\beta} \delta \phi \right] + {\cal O}(r^{-3}) \; .
\label{eqn:ThetaasymE}
\end{eqnarray}
where
\begin{equation}
\overline{v}^{\beta} \equiv \frac{\partial \overline{f}}{\partial 
(\overline{\nabla}_{\beta} \phi)} \; .
\label{eqn:vbardef}
\end{equation}
We see from 
Eqs.(\ref{eqn:QasymEo}) and (\ref{eqn:ThetaasymE}) that the 
contributions, $H_{\mu \nu}^{~~\rho}$ and $\delta g^{\mu \nu}$, of the 
exotic matter, $g^{\mu \nu}$, to the mass and the angular 
momentum vanish. The relation between $\overline{v}^{\beta}$ and 
$v^{\beta}$ is obtained, from Eq.(\ref{eqn:vbardef}), as
\begin{eqnarray}
\overline{v}^{\beta} & = & \frac{\partial \overline{f}}{\partial 
(\nabla_{\beta} \phi)} \nonumber \\
& = & \overline{g}^{\mu \nu} \frac{\partial R_{\mu \nu}}{\partial 
(\nabla_{\beta} \phi)} - \frac{\sqrt{- g}}{\sqrt{- \overline{g}}} \left[ 
\frac{\partial f}{\partial R_{\mu \nu}} 
\frac{\partial R_{\mu \nu}}{\partial (\nabla_{\beta} \phi)} - 
\frac{\partial f}{\partial 
(\nabla_{\beta} \phi)} \right] \nonumber \\
& = & \frac{\sqrt{- g}}{\sqrt{- \overline{g}}} \: v^{\beta} \; ,
\end{eqnarray}
where $R_{\mu \nu}$ is given by Eq.(\ref{eqn:Riccisolved}), and 
Eq.(\ref{eqn:gbardef}) is used.
So, $\overline{v}^{\beta} \delta \phi \rightarrow \delta C^{\beta} + 
{\cal O}(r^{-3})$, and we find $\overline{B}^{\beta}$ such that 
$\overline{\Theta}^{\beta} \rightarrow \delta \overline{B}^{\beta} + 
{\cal O}(r^{- 3})$, which is given by 
\begin{eqnarray}
\overline{B}^{\beta} & = & \frac{1}{16 \pi} \left[ g^{\: -1}_{\mu \nu} 
g^{\rho \beta} 
\partial_{\rho} 
\overline{g}^{\mu \nu} - \partial_{\nu} \overline{g}^{\nu \beta} + 
C^{\beta} \right] + {\cal O}(r^{-3}) \nonumber \\
& = & \frac{1}{16 \pi} \left[ g^{\: -1}_{\mu \nu} g^{\rho \beta} 
\partial_{\rho} g^{\mu \nu} 
- \partial_{\nu} g^{\nu \beta} + C^{\beta} - \left( \nabla_{\nu} 
P^{\nu \beta} + \frac{1}{2} \nabla^{\beta} P \right) \right] + 
{\cal O}(r^{-3})
\nonumber \\
& = & B^{\beta} - \frac{1}{16 \pi} \left( \nabla_{\nu} P^{\nu \beta} + 
\frac{1}{2} 
\nabla^{\beta} P \right) + {\cal O}(r^{-3}) \; ,
\label{eqn:BasymE}
\end{eqnarray}
where we have rewritten the equation in terms of the tensors in the 
original frame, by using Eqs.(\ref{eqn:BO}) and (\ref{eqn:gbardef2}) 
with the facts $P^{\mu \nu} \rightarrow g^{\mu \nu} + {\cal O}(r^{- 1})$ 
and thus 
$\partial_{\rho} P^{\mu \nu} \sim {\cal O}(r^{- 2})$. 

Then, from 
Eqs.(\ref{eqn:QasymE}) and (\ref{eqn:BasymE}), we obtain 
Eq.(\ref{eqn:IntegrandInf}) with $\chi^{\mu}$ being 
$\xi^{\mu}_{(t)}$ and $\xi^{\mu}_{(\varphi)}$. 
Therefore, the mass and the angular momentum are the same 
between the original frame and the Einstein frame. 

\section{Conclusion and Discussion}

In summary, we considered a generalized theory of gravity whose 
Lagrangian is an arbitrary function of the metric, the Ricci tensor, a 
scalar field and its derivative, which is converted into the 
Einstein frame via a ``Legendre" transformation or a conformal 
transformation. By following the 
definition of the thermodynamical variables formulated in 
Ref.\cite{Wald,IyerWald}, we have shown that all the thermodynamical 
variables are the same between the original frame and the Einstein 
frame, under the assumptions that the spacetimes in both frames are 
asymptotically flat, regular and possess event horizons with 
non-zero temperatures. This conclusion is based on 
the equivalence of the structure of the horizon, and of the potential, 
Eq.(\ref{eqn:potofomega}), of the symplectic current 3-form, 
$\omega_{\mu \nu \rho}$, both at the horizon and in the 
asymptotically flat region.

The above result is a generalization of the works by Jacobson, Kang and 
Myers\cite{JKM1,JKM2}. They showed the entropy, defined in 
Ref.\cite{Wald,IyerWald}, in the theory whose 
Lagrangian is given by either
\begin{equation}
I = \frac{1}{16 \pi} \int d^4 x \: \sqrt{- g} \: f(R, \: \phi, 
\nabla_{\alpha} \phi)
\label{eqn:SimpleAction}
\end{equation}
or
\begin{equation}
I = \frac{1}{16 \pi} \int d^4 x \: \sqrt{- g} \: \left[ R + \alpha R^2 + 
\beta R_{\mu \nu} R^{\mu \nu} \right]
\end{equation}
coincides with that in the Einstein frame. In the latter case, they used 
the fact in their proof that the asymptotic structure in the Einstein 
frame, such as the mass 
and the angular momentum, is the same as that in 
the original frame. However, if we consider a scalar field 
non-minimally coupled to gravity, which appears, e.g., in the 
Brans--Dicke 
theory and the effective string theory, the scalar field 
could change the asymptotic structure, in general. Actually, there is 
an example\cite{Will} where the gravitational mass in the Einstein 
frame is different from that in the Brans--Dicke frame. We have 
proved, even in such cases, the ``thermodynamical" mass defined by 
Eq.(\ref{eqn:Mdef}), as well as the angular momentum defined by 
Eq.(\ref{eqn:Jbdef}), are the same between the two frames, by showing 
the 
equality of the potential of the symplectic current 3-form in the 
asymptotically flat region. Thus, the definition of 
Ref.\cite{Wald,IyerWald} is consistent with the field redefinition 
method\cite{JKM1,JKM2}, not only for the entropy, but also the mass 
and the angular momentum. 

We have not considered, in the action(\ref{eqn:actionorign}), the higher 
order derivatives of the scalar field or other matter fields, such as 
gauge fields\cite{SudarskyWald} and an axion field, for simplicity. 
However, we expect that the conclusion does not change even if we 
include 
such terms, because matter fields do not contribute to the black hole 
entropy directly\cite
{IyerWald,JKM2} and it can be shown that their contributions to the 
mass and the angular momentum are the same in the two frames, by a 
similar discussion to that in the previous section. Our results seem 
to be extended into the higher dimension straightforwardly, although 
here we discussed only 4-dimensional spacetimes. The 
generalization into the theories whose Lagrangian includes the Weyl 
tensor 
and the derivatives of the Riemann tensor also seems to be interesting. 
While a ``Legendre" transformation in such cases is  discussed by 
Magnano 
et al.\cite{Magnano2}, however, the action is not transformed into the 
Einstein frame in those cases. Therefore, the possibility of such 
generalization is not clear at present. 

In Ref.\cite{TMT1}, stability of non-Abelian black hole solutions in the 
Brans--Dicke theory is considered, using catastrophe theory. It is 
claimed that we can understand the stability of such black holes 
naturally when we choose, as the catastrophe variables, the entropy and 
the gravitational mass in the Einstein frame, which is equal to the ADM 
mass and also to the ``thermodynamical'' mass, whereas choice of the 
area and the gravitational mass in the Brans--Dicke frame is not 
appropriate for discussing the stability of the black holes by 
catastrophe theory. Stability of those black holes is analyzed also by 
the linear perturbation method, which gives the same result as 
catastrophe theory\cite{TMT2}. Since the entropy and the gravitational 
mass in the Einstein frame are the same as the entropy and the 
``thermodynamical" mass in the Brans--Dicke frame, respectively, 
the above results suggest that the thermodynamical variables 
defined in 
Ref.\cite{Wald,IyerWald} are deeply related to the dynamical stability 
of the black 
holes. Black holes in a theory with a dilaton and its coupling to the 
Gauss--Bonnet combination of curvature invariants are also analyzed in 
the same context with a similar conclusion\cite{TM}. In addition, the 
second law is satisfied for quasi-stationary processes where 
perturbations of infalling matter, without the effects of the higher 
curvature interactions, are considered on a stationary black 
hole\cite{JKM1}. All these suggest that the thermodynamical variables, 
which are defined formally as described in Section III, do have physical 
meanings at least in the case of stationary black holes. 

In dynamical processes of black holes, however, the definition of the 
entropy is still ambiguous because of the absence of the bifurcation 
surface, and 
hence the second law has not been established in general. One example 
which satisfies the second law is the case where the action is given 
by 
Eq.(\ref{eqn:SimpleAction})\cite{JKM1,Kang}. In this case, the theory is 
converted, via a conformal transformation, into Einstein gravity 
with scalar fields (one scalar field when the Lagrangian is linear in the 
scalar curvature, and two fields otherwise). Then, showing that the null 
energy 
condition in the Einstein frame is satisfied, we can prove that the 
entropy in the 
Einstein frame, which is proportional to the area, is a non-decreasing 
quantity, due to Hawking's area theorem in Einstein gravity. 
Therefore, the second law holds, if we define the entropy in the original 
frame to be equal to that in the Einstein frame also in the dynamical 
processes. This definition of the dynamical entropy differs 
from that 
proposed in Ref.\cite{IyerWald}, which uses boost invariant parts of the 
field variables, but the second law has not been 
established for the latter definition even in the case of the 
action(\ref{eqn:SimpleAction}). In general, 
as the above example, the 
equivalence of the entropy between the original frame and the Einstein 
frame for stationary black holes will provide a useful clue to 
analyze the dynamical entropy, since useful results have been already 
established in Einstein gravity. 

It does not seem to be easily accomplished to define the dynamical 
entropy in the 
general case of the 
action(\ref{eqn:actionorign}), because a null vector in the original 
frame may be transformed to either a timelike, null or spacelike vector 
in the Einstein frame, and the energy condition of the exotic matter, 
$g^{\mu \nu}$, in the Einstein frame is not trivial. Therefore, 
investigations 
of the second law in arbitrary quasi-stationary processes, which 
include the effects of the higher curvature interactions, seem to be 
important. Our result can be applied to such analyses, since the effects 
of the higher curvature interactions are included in the 
action(\ref{eqn:actionEinstein}) in the Einstein frame. We may clarify 
under what condition the second law is satisfied, or whether we need 
to 
consider the generalized second law, at least in the quasi-stationary 
processes, which might afford us a facility in the analysis of the 
dynamical entropy, i.e., its general definition and the second law.

\vspace{0.5cm}
{\bf Acknowledgment}

We would like to thank J. Barrow, G. Gibbons, G. 
Kang, T. Tamaki and T. Torii for useful discussions, 
T. Jacobson for calling our attention to the 
papers by him and his collaborators, 
and P. Haines for 
reading the manuscript. 
K. M. also acknowledges J. Barrow for his kind hospitality at
University of Sussex, where this work was completed.
This work was supported partially by the
Grant-in-Aid for Scientific Research  Fund of the
Ministry of Education, Science and Culture  (No.
08102010), and by Waseda University
Grant for Special Research Projects.


\appendix
\section{Derivation of $\Theta^{\beta}$ and $Q^{\beta \alpha}$}

In this appendix, we derive the explicit forms of 
$\Theta^{\beta}$ and $Q^{\beta \alpha}$ both in the original frame and 
in the Einstein frame. 

We begin with an action given by 
\begin{equation}
I = \frac{1}{16 \pi} \int d^4 x \: \sqrt{- g} \; L(g^{\alpha \beta}, 
R_{\alpha \beta \rho \lambda}, \phi, \nabla_{\alpha} \phi, s^{\alpha 
\beta}, \nabla_{\rho} 
s^{\alpha \beta}) \: ,
\end{equation}
where $L$ is a scalar function of the metric, $g^{\alpha \beta}$, the 
Riemann tensor, $R_{\alpha \beta \rho \lambda}$, a scalar field, 
$\phi$, a symmetric second rank tensor, $s^{\alpha \beta}$, and the 
derivatives of $\phi$ and $s^{\alpha \beta}$. Variation of this action 
gives, for 
the equation of motion, $E^{(i)}$, of $\psi_{(i)}$ and $\Theta^{\beta}$, 
as 
\begin{eqnarray}
E^{(g)}_{~ \mu \nu} & = & \frac{1}{16 \pi} \biggl[ M_{\mu \nu} - 
\frac{1}{2} g_{\mu \nu} L - X^{\alpha \beta \rho}_{~~~~ ( \mu}  R_{\nu ) 
\rho \beta \alpha}  - 2 \nabla_{\rho} \nabla_{\lambda} X^{~~ \lambda 
\rho}_{( \mu ~~ \nu )}  \biggr. \nonumber \\
& & \biggl. + \nabla_{\alpha} \left( K_{\beta ( \nu \mu )}
s^{\alpha \beta} + K_{\beta ( \nu}^{~~~ \alpha} s^{\beta}_{~ \mu )} - 
K_{\beta ~ ( \nu}^{~ \alpha} s^{\beta}_{~ \mu )} \right) \biggr] \; ,
\label{eqn:gnEqofg} \\
E^{(s)}_{~\mu \nu} & = & \frac{1}{16 \pi} \left[ 
N_{\mu \nu} - 
\nabla_{\rho} 
K_{\mu \nu}^{~~\rho} \right] \; ,
\label{eqn:gnEqofs} \\
E^{(\phi)} & = & \frac{1}{16 \pi} \left[ \frac{\partial L}{\partial \phi} - 
\nabla_{\mu} w^{\mu} \right] \; ,
\label{eqn:gnEqofphi} \\
\Theta^{\beta} & = & \frac{1}{16 \pi} \biggl[ 2 \left( 
\nabla_{\alpha} X^{~ ( \alpha \beta )}_{\mu ~~~ \: \nu} \right) \delta 
g^{\mu \nu} 
- 2 X^{~ ( \alpha \beta)}_{\mu ~~~ \: \nu} \nabla_{\alpha} \delta 
g^{\mu \nu} 
\biggr.  \nonumber \\
& & \biggl. + K_{\alpha ~ \mu}^{~ \beta} s^{\alpha}_{~ \nu} \delta 
g^{\mu \nu} - 
K_{\alpha \nu \mu} s^{\alpha \beta} \delta g^{\mu \nu} - 
K_{\alpha \nu}^{~~ \beta} s^{\alpha}_{~ \mu} \delta 
g^{\mu \nu} + w^{\beta} \delta \phi + K_{\mu \nu}^{~~\beta} \delta 
s^{\mu \nu} \biggr] \: ,
\label{eqn:gnTheta}
\end{eqnarray}
where 
\begin{eqnarray}
M_{\mu \nu} & \equiv & \frac{\partial L}{\partial g^{\mu \nu}} \; , 
~~~~~~
X^{\mu \nu \rho \lambda} \; \equiv \; \frac{\partial L}{\partial 
R_{\mu \nu \rho \lambda}} \; , ~~~~~~
K_{\mu \nu}^{~~\rho} \; \equiv \; \frac{\partial L}{\partial \left( 
\nabla_{\rho} s^{\mu \nu} \right)} \; , \nonumber \\
N_{\mu \nu} & \equiv & \frac{\partial L}{\partial s^{\mu \nu}} \; , 
~~~~~~
w^{\mu} \; \equiv \; \frac{\partial L}{\partial \left( \nabla_{\mu} \phi 
\right)} \; .
\end{eqnarray}
Then we regard the variation as a coordinate transformation induced by 
an arbitrary vector, $\zeta^{\mu}$, and obtain 
$J^{\beta} = \Theta^{\beta} - \zeta^{\beta} L$ of the form
\begin{eqnarray}
J^{\beta} & = & \frac{1}{8 \pi} \biggl[ \nabla_{\alpha} \left\{ X^{\alpha 
\beta \mu \nu} \nabla_{\mu} \zeta_{\nu} - 2 \zeta_{\nu} \nabla_{\mu} 
X^{\alpha \beta \mu \nu} + \zeta^{\nu} K_{\mu ~ \: \nu}^{~ [ \alpha} 
s^{\beta ] \mu} 
+ \zeta^{\nu} K_{\mu}^{~ [ \alpha \beta ] } s^{\mu}_{~ \nu} + 
\zeta^{\nu} K_{\mu \nu}^{~~ [ \alpha} s^{\beta ] \mu} \right\} \biggr. 
\nonumber \\ 
& & 
+ E^{(g) \beta}_{~ \mu} \zeta^{\mu} + E^{(s)}_{~ \alpha \mu} 
s^{\beta \alpha} \zeta^{\mu} - 2 \zeta^{\mu} \left(
M_{\mu}^{~ \beta} - 2 R_{\alpha \nu \rho \mu} X^{\alpha \nu \rho \beta} 
- \frac{1}{2} w^{\beta} \nabla_{\mu} \phi \right. 
\nonumber \\
& & \biggl. \left.
- \frac{1}{2} 
K_{\alpha \nu}^{~~ \beta} \nabla_{\mu} s^{\alpha \nu} + 
K_{\alpha \mu}^{~~ \rho} \nabla_{\rho} s^{\alpha \beta} + N_{\mu 
\alpha} s^{\alpha \beta} \right) \biggr] \: ,
\label{eqn:Jrewritten}
\end{eqnarray}
by integrating by parts and substituting Eqs.(\ref{eqn:gnEqofg}) and 
(\ref{eqn:gnEqofs}). 
Since $L$ is a scalar, we have 
\begin{eqnarray}
0 & = & {\cal L}_{\eta} L - \eta^{\mu} \nabla_{\mu} L \nonumber \\
& = & - 2 
\left[ M_{\mu}^{~ \beta} - 2 R_{\alpha \nu \rho \mu} X^{\alpha \nu \rho 
\beta} - \frac{1}{2} w^{\beta} \nabla_{\mu} \phi \right.
\nonumber \\
& & - \left.
\frac{1}{2} K_{\alpha \nu}^{~~ \beta} \nabla_{\mu} s^{\alpha \nu} + 
K_{\alpha \mu}^{~~ \rho} \nabla_{\rho} s^{\alpha \beta} + N_{\mu 
\alpha} s^{\alpha \beta} \right] \left( \nabla_{\beta} 
\eta^{\mu} \right) \: ,
\end{eqnarray}
for an arbitrary vector, $\eta^{\mu}$, and thus the inside of the round 
bracket in Eq.(\ref{eqn:Jrewritten}) vanishes identically. Then the 
potential $Q^{\beta \alpha}$ of $J^{\beta}$ is given as 
\begin{equation}
Q^{\beta \alpha} = \frac{1}{8 \pi} \left\{ X^{\alpha \beta \mu \nu} 
\nabla_{\mu} \zeta_{\nu} - 2 \zeta_{\nu} \nabla_{\mu} X^{\alpha \beta 
\mu \nu} + \zeta^{\nu} K_{\mu ~ \: \nu}^{~ [ \alpha} s^{\beta ] \mu} + 
\zeta^{\nu} K_{\mu}^{~ [ \alpha \beta ]} 
s^{\mu}_{~ \nu} + \zeta^{\nu} K_{\mu \nu}^{~~ [ \alpha} s^{\beta ] \mu} 
\right\} \: , 
\label{eqn:gnQ}
\end{equation}
when the equations of motion, $E^{(g)}_{~ \mu \nu} = 0$ and 
$E^{(s)}_{~ \mu \nu} = 0$, are satisfied.

In the original frame, $X^{\mu \nu \rho \lambda}$, $w^{\mu}$ and 
$K_{\mu \nu}^{~~ \rho}$ are given as 
\begin{equation}
X^{\mu \nu \rho \lambda} = \frac{1}{2} \left( 
P^{\nu [ \lambda} 
g^{\rho ] \mu} + P^{\mu [ \rho} g^{\lambda ] \nu} \right) \; , ~~~~~ 
w^{\mu} = v^{\mu} \; , ~~~~~ 
K_{\mu \nu}^{~~ \rho} = 0 \; . 
\label{eqn:varisO}
\end{equation}
Then, by substituting Eq.(\ref{eqn:varisO}) into 
Eqs.(\ref{eqn:gnTheta}) and (\ref{eqn:gnQ}), $\Theta^{\beta}$ and 
$Q^{\beta \alpha}$ in the original frame are calculated as
\begin{eqnarray}
\Theta^{\beta} & = & \frac{1}{32 \pi} \left[ \left( \nabla_{\nu} 
P^{\beta}_{~ \mu} \right) \delta g^{\mu \nu} - g^{\: -1}_{\mu \nu} \left( 
\nabla_{\alpha} P^{\alpha \beta} \right) \delta g^{\mu \nu} + \left( 
\nabla_{\mu} P^{\mu}_{~ \alpha} \right) \delta g^{\alpha \beta}  \right. 
\nonumber \\
& & - \left( \nabla^{\beta} P_{\mu \nu} \right) \delta g^{\mu \nu} 
+ g^{\: -1}_{\mu \nu} P^{\alpha \beta} \nabla_{\alpha} \delta 
g^{\mu \nu} - 
P^{\nu}_{~ \alpha} \nabla_{\nu} \delta g^{\alpha \beta}  \nonumber \\
& & \left. - P^{\beta}_{~ \nu} \nabla_{\mu} \delta 
g^{\mu \nu} + 
P_{\mu \nu} \nabla^{\beta} \delta g^{\mu \nu} + 2 v^{\beta} \delta \phi 
\right] \label{eqn:AThetaO} \; , \\
Q^{\beta \alpha} & = & \frac{1}{16 \pi} \left[ P^{\nu [ \alpha} 
\nabla_{\nu} \zeta^{\beta ]} + P^{\nu [ \beta} 
\nabla^{\alpha ]} \zeta_{\nu} + 2 \left( \zeta_{\nu} \nabla^{[ \beta} 
P^{\alpha ] \nu} + \zeta^{[ \alpha} 
\nabla_{\nu} P^{\beta ] \nu} \right) \right] \label{eqn:AQO} \; .
\end{eqnarray}

On the other hand, in the Einstein frame, we have
\begin{equation}
X^{\mu \nu \rho \lambda} = 
\overline{g}^{\nu [ \lambda} \overline{g}^{\rho ] \mu} \; , ~~~~~ 
w^{\mu} = \overline{v}^{\mu} \; , ~~~~~ 
K_{\mu \nu}^{~~ \rho} = H_{\mu \nu}^{~~ \rho} \: ,
\label{eqn:varisE}
\end{equation}
and hence, we obtain $\overline{\Theta}^{\beta}$ and 
$\overline{Q}^{\beta \alpha}$ in the Einstein frame, as
\begin{eqnarray}
\overline{\Theta}^{\beta} & = & \frac{1}{16 \pi} \biggl[ 
\overline{g}^{\: -1}_{\mu \nu} \overline{\nabla}^{\beta} \delta 
\overline{g}^{\mu \nu} - \overline{\nabla}_{\nu} \delta 
\overline{g}^{\nu 
\beta} + H_{\mu \nu}^{~~\beta} \delta g^{\mu \nu} + \overline{v}^{\beta} 
\delta \phi \biggr. \nonumber \\
& & + \biggl. 
H_{\mu ~ \alpha}^{~ \beta} g^{\mu}_{~ \nu} \delta 
\overline{g}^{\alpha \nu} - H_{\mu \nu \alpha} g^{\beta \mu} \delta 
\overline{g}^{\alpha \nu} - H_{\mu \nu}^{~~ \beta} g^{\mu}_{~ \alpha} 
\delta 
\overline{g}^{\alpha \nu}  \biggr] \; , \label{eqn:AThetaE} \\
\overline{Q}^{\beta \alpha} & = & \frac{1}{8 \pi}  \left[ 
\overline{\nabla}^{[ \alpha} \zeta^{\beta ]} + \zeta^{\nu} 
H_{\mu ~ \: \nu}^{~ [ \alpha} g^{\beta ] \mu} +  \zeta^{\nu} 
H_{\mu}^{~ [ \alpha \beta ]} g^{\mu}_{~ \nu} 
 + \zeta^{\nu} H_{\mu \nu}^{~~ [ \alpha} g^{\beta ] \mu} \right] 
\label{eqn:AQE} \; .
\end{eqnarray}

\vskip 2cm
\baselineskip .2in



\begin{thebibliography}{99}
\bibitem{NoteDet} The matrix $\partial^2 f / \partial R_{\mu \nu} 
\partial R_{\rho \sigma}$ is regarded as a $10 \times 10$ matrix, 
whose elements are 
given by the second derivative of $f$ by the independent components of 
$R_{\mu \nu}$\cite{Magnano2}.
\bibitem{NoteTotDiv} The thermodynamical variables do not change 
even if we include the totally divergent term in the case of stationary 
black holes. However, it results in ambiguity of the definition of the 
entropy in dynamical processes\cite{IyerWald}.
\bibitem{note2}  The essential part of this subsection has been 
already noted in Ref.\cite{JKM2}.
\end{thebibliography}
\end{document}